# The role of silver addition on the structural and superconducting properties of polycrystalline $Sr_{0.6}K_{0.4}Fe_2As_2$

Lei Wang, Yanpeng Qi, Zhaoshun Gao, Dongliang Wang,

Xianping Zhang, Yanwei Ma[*]

Key Laboratory of Applied Superconductivity, Institute of Electrical Engineering,

Chinese Academy of Sciences, P. O. Box 2703, Beijing 100190, China

**Abstract**

The effect of Ag addition (0-20 wt%) on polycrystalline $Sr_{0.6}K_{0.4}Fe_2As_2$ superconductor has been investigated. It is found that the critical transition temperature $T_c$ was not depressed, and the irreversibility field $H_{irr}$ and hysteresis magnetization were significantly enhanced upon Ag addition. Characterization study reveals that larger grains are observed in the Ag-added samples. Moreover, the formation of glassy phase as well as amorphous layer, which are present in almost all the grain edges and boundaries in pure samples, are suppressed by Ag addition. The improvement of superconducting properties in Ag-added samples may originate from the enlargement of grains as well as better connections between grains

[*] Author to whom correspondence should be addressed; E-mail: ywma@mail.iee.ac.cn

## Introduction

Feverish activity on superconductivity in iron pnictide materials has followed the discovery of superconductivity at 26 K in the LaFeAsO$_{1-x}$F$_x$ compound [1], due to the high critical temperature $T_c$ and underlying mechanism. Critical temperatures up to 55 K have been achieved in fluoride doped SmFeAs(O$_{1-x}$F$_x$) [2]. The related iron-based superconductor A$_{1-x}$K$_x$Fe$_2$As$_2$ (A=Ba and Sr) also exhibits a $T_c$ of ~38 K [3-5]. These recent discovered iron pnictide superconductors have been accepted as the second class of high-$T_c$ superconductors. Like YBa$_2$Cu$_3$O$_{7-\delta}$, the iron pnictide superconductors show high critical current density $J_c$, high upper critical field $H_{c2}$ and irreversibility field $H_{irr}$, low anisotropy $\gamma = H_{c2}^{(ab)} / H_{c2}^{(c)}$ [6-11]. Very recently, intrinsic weak link behavior, similar to high-$T_c$ cuprates, has been proved in thin film Ba(Fe$_{1-x}$Co$_x$)$_2$As$_2$ bicrystals [12], suggesting that texture fabrication process is needed for their potential applications.

The introduction of silver into high-$T_c$ superconductors is known to improve their structural and superconducting properties [13-16]. It was reported that silver promotes the c-axis orientation and the crystallization of the superconducting phase, and catalyzes the intergranular coupling of the superconducting grains in Bi-(Pb)-Sr-Ca-Cu-O and YBa$_2$Cu$_3$O$_y$ ceramics [13, 14]. An unusual $J_c$ enhancement has been observed in Ag added YBa$_2$Cu$_4$O$_8$ high-temperature superconductor [15]. Another study revealed that the beneficial effect of Ag addition on superconducting properties originates from the suppression of the formation of liquid phase in the grain boundaries by silver [16]. However, the effect of Ag addition on the new iron-based superconductors, such as Sr$_{0.6}$K$_{0.4}$Fe$_2$As$_2$, has not been reported.

In this paper, we introduce silver by mixing metallic Ag powder with well ground Sr$_{0.6}$K$_{0.4}$Fe$_2$As$_2$ raw materials. Resistance, magnetic susceptibility and hysteresis of the pure and Ag added polycrystalline Sr$_{0.6}$K$_{0.4}$Fe$_2$As$_2$ have been measured. Results of structural study through scanning electron microscope (SEM) and transmission electron microscope (TEM) are also presented.

## Experimental details

The polycrystalline samples investigated were prepared by a one-step method developed by our group [17] together with ball milling process. Sr filings, Fe powder, As and K pieces, with a ratio Sr : K : Fe : As = 0.6 : 0.44 : 2 : 2.2, were thoroughly ground in Ar atmosphere for more than 10 hours using ball milling method. For the reasons of simplicity, in the following the nominal composition will be described as $Sr_{0.6}K_{0.4}Fe_2As_2$. After the ball milling, the raw powders were divided into three parts, two of which were added with various amount of Ag powder (5 wt% and 20 wt%). Then, the three kinds of powder were ground in a mortar for half an hour, respectively. The final powders were encased and sealed into Nb tubes (OD: 8 mm, ID: 5 mm), sintered at 500 ℃ for 15 hours, heated to around 900 ℃ in 5 hours and then kept for 35 hours in Ar atmosphere. The density of the sintered sample is about 70% of the theoretical value of 5.89 g·cm$^{-3}$.

Phase identification was characterized by powder X-ray diffraction (XRD) analysis with Cu-Kα radiation from 10 to 80˚. Resistivity measurements were carried out by the standard four-probe method using a PPMS system. Magnetization measurements were performed with a PPMS system in fields up to 7 T. Magnetic critical current densities were calculated using Bean model. Microstructural observations were performed using scanning electron microscope (SEM) and transmission electron microscope (TEM). Our TEM specimens were prepared as follow: $Sr_{0.6}K_{0.4}Fe_2As_2$ powder was scraped from a sintered bulk. The powder was dispersed in suitable absolute ethanol using an ultrasonic-wave machine. Then, a drop of the suspension was placed on holey carbon coated copper grid, which was subsequently dried.

**Results and discussion**

Figure 1 shows X-ray diffraction (XRD) patterns for the pure and Ag added polycrystalline $Sr_{0.6}K_{0.4}Fe_2As_2$. The XRD pattern for the pure sample reveals an almost single phase, $Sr_{0.6}K_{0.4}Fe_2As_2$. The Ag added samples consist of $Sr_{0.6}K_{0.4}Fe_2As_2$ as the major phase, with some Ag and a small amount of impurity phases, which were identified as FeAs and AgSrAs. Clearly, the addition of Ag induces small amount of impurity phases without destroying the parent compound.

The temperature dependences of resistivity for various samples are shown in Fig.

2. Resistivity drops at 35 K, and vanishes at about 33 K for all samples. The Ag added samples show a slightly sharpened transition, compared to the pure sample. The inset of Fig. 2 shows temperature dependences of the zero-field cooled (ZFC) and field cooled (FC) magnetic susceptibility of the pure and Ag added samples. Below 34 K, the ZFC signal is diamagnetic for all samples, which is consistent with the resistive results. From this, one can also roughly estimate the shielding fraction. The shielding fraction at 5 K is 83 % and 90 % for the pure and 5% Ag added sample, respectively. One can see that the temperature dependence of dc shielding fraction is improved by 5% Ag addition, and 87 % is achieved at 25 K. The ZFC curve for the 20% Ad added sample shows a slightly decreased shielding fraction at 5 K, because of a large volume fraction of Ag. It clearly proves that the Ag addition leads to an improvement of uniformity in superconducting properties.

To obtain information about the upper critical field $H_{c2}$ and irreversibility field $H_{irr}$, the variation of resistivity under various magnetic fields (H=0, 1, 3, 5, 7 and 9 T) was studied. As the field is increased, the onset critical transition temperature for all samples shifts with magnetic field weakly. But the zero resistance temperature of the pure sample decreases more rapidly than that of the Ag added samples. Using criteria of 90% and 10 % of normal state resistivity, upper critical field $H_{c2}$ ($T$) and irreversibility field $H_{irr}$ ($T$) can be estimated, respectively. The resultant $H$-$T$ phase diagrams for the samples are shown in figure 3. $H_{c2}$ ($T$) was not significantly changed by Ag addition. The upper critical field at zero-temperature $H_{c2}$ (0) was calculated using the Werthamer-Helfand-Hohenberg (WHH) formula, $H_{c2}$ (0) = - 0.693 $T_c$ (d$H_{c2}$/d$T$). The slope d$H_{c2}$/d$T$ is estimated from the $H$-$T$ phase diagram，and the value is about 7.9 for these samples. Taking $T_c$ = 34 K, the upper critical field is $H_{c2}$ (0) = 186 T, which is consistent with the $H_{c2}^{ab}$ (0) = 185.4 T reported in single crystal [9].

In fact, applications are limited by a lower characteristic field, the irreversibility field $H_{irr}$ ($T$) at which $J_c$ vanishes. The prominent feature of irreversibility fields is that the $H_{irr}$-$T$ curve becomes steeper as a result of Ag addition. In particular, the $H_{irr}$ value for the pure sample is 3 T at 30 K, however, the extrapolated $H_{irr}$ for the 5 % Ag added sample at 30 K is about 15 T, 12 T higher than that of the pure sample.

Magnetic hysteresis measurements have been carried out on rectangular specimens in the same dimension of 3×2×1 mm$^3$, which were cut from the bulks. Critical current density ($J_c$) derived from the magnetic hysteresis for various samples at 5 and 20K is shown in Figure 4. As can be seen, the Jc increased greatly with increasing Ag addition content. The $J_c$ of 20 % Ag added samples at 5 K in self field is about 2.5×10$^4$ A/cm$^2$ and remains above 1.5×10$^3$ A/cm$^2$ beyond 6.5 T, twice as high as for the pure sample. Most importantly, excellent $J_c$ of about 1.0×10$^4$ A/cm$^2$ at 20 K was achieved in the 20% Ag added samples (see the inset).

The microstructure of the samples was studied using a scanning electron microscope. Figure 5 show scanning electron micrographs of the pure, 5 % and 20 % Ag added samples. The pure and 5 % Ag added samples consist of small grains, with an average size of 0.5-2μm, which was identified as $Sr_{0.6}K_{0.4}FeAs$ using EDS. These grains are much smaller than those reported previously [18], due to the ball milling process used in our experiments. However, some large grains appear in the 20 % Ag added sample, and it seems that Ag addition have a large effect on grain growth. These large grains, meaning large dimensions of intra-grain loops, were supposed to contribute to the enhancements of magnetic $J_c$. Some Ag particles can be seen in the Ag added samples, as shown in Fig. 5d by quadrant back scattering detector.

Further study on the microstructure of the samples was performed using transmission electron microscopy. Figure 6a show a typical grain of the pure sample. Glassy phase (wetting phase or liquid phase) around the grain was found. Detailed observation of a single grain was performed, and an amorphous layer of several nanometers in thickness around individual grains was observed in the pure sample (Fig. 6b). Evidence of similar glassy phase and amorphous layer was also observed in polycrystalline $REFeAsO_{1-x}$ and $YBa_2Cu_3O_{7-\delta}$ [19, 20, 16]. Numerous randomly selected grains have been examined in this study, and it is confirmed that the amorphous layer is ubiquitous in the pure bulk. The amorphous layer was also found in at the interface between two grains, as shown in Fig. 6c. In some cases, a particle consists of several small grains. The boundaries between the small grains were also studied. However, more information is hindered by the glassy phase and amorphous layer on the particle (Fig. 6d). Although clear grain boundaries cannot yet be ruled out, the super-current across the grain boundaries can be largely reduced by the glassy phase and amorphous layer [19, 20].

By contrast, TEM study on the Ag added samples reveals almost no glassy phase around individual grains, as shown in Fig. 7a. Except for similar amorphous layers, numerous clean grain edges, in various crystallographic directions, were found in Ag added sample. A typical clean grain edges is shown in Fig. 7b. Grain boundaries between two grains can be studied in details (Fig. 7c), because of the absence of glassy phase around grains and particles. It can be seen that, the part of boundary near the edge of grains is of an amorphous layer of about 4 nm in thickness, whereas, the inner part of boundary is clean, with some dislocations. Observations of clean boundaries were also performed (Fig. 7d). It reveals that, the connections of grains are perfect. Two kinds of grain boundary were seen, one being a clear plane with a small distortion (Indicated by white dashed line), and the other a large amount of dislocations. Clearly, the glassy phase and amorphous layer were suppressed, and better connections between grains formed in Ag added samples.

The effect of Ag addition on polycrystalline $Sr_{0.6}K_{0.4}Fe_2As_2$ is much similar to that on $YBa_2Cu_3O_{7-\delta}$ [16]. Ag addition has a large effect on the grain growth and glassy phase formation, as supported by SEM and TEM observations. Based on microstructural analyses and magnetization data, we may say that larger grains and good grain connection caused by Ag addition are responsible for the better performance in our Ag-added samples. Actually, through Ag addition, we have successfully observed large transport critical currents in Fe-based superconducting wires and tapes, for details, please see the reference [21]

**Conclusions**

Critical transition temperature, upper critical field, irreversibility field and critical current density $J_c$ have been studied in order to understand the role of silver addition on $Sr_{0.6}K_{0.4}Fe_2As_2$ superconductor. The addition of silver up to 20 wt% does not depress the critical transition temperature. Significant enhancement of irreversibility field $H_{irr}$ and critical current density $J_c$ have been observed for the Ag added samples. Results of structural study suggest that, Ag addition has a large effect on the grain growth and glassy phase formation, and the enhancement of superconducting properties is attributed to the enlargement of grains as well as better connections between grains.

**Acknowledgements**

The authors thank Profs. Xuedong Bai, Haihu Wen, Liye Xiao and Liangzhen Lin for their help and useful discussions. This work was partially supported by the Beijing Municipal Science and Technology Commission under Grant No. Z09010300820907, National Science Foundation of China (grant no. 50802093) and the National '973' Program (grant no. 2006CB601004).

# References

1. Y. Kamihara, T. Watanabe, M. Hirano and H. Hosono, *J. Am. Chem. Soc.* **130**, 3296 (2008).

2. Z. A. Ren, W. Lu, J. Yang, W. Yi, X. L. Shen, Z. C. Li, G. C. Che, X. L. Don, L. L. Sun, F. Zhou, Z. X. Zhao, *Chinese Phys. Lett.* **25**, 2215-2216 (2008)

3. M. Rotter, M. Tegel, D. Johrendt, *Phys. Rev. Lett*. **101**, 107006 (2008)

4. G. F. Chen, Z. Li, G. Li, W. Z. Hu, J. Dong, X. D. Zhang, P. Zheng, N. L. Wang, J. L. Luo, *Chin. Phys. Lett.* **25**, 3403 (2008)

5. K. Sasmal, B. Lv, B. Lorenz, A. Guloy, F. Chen, Y. Xue, C. W. Chu, *Phys. Rev. Lett*. **101**, 107007 (2008)

6. A. Yamamoto, J. Jaroszynski, C. Tarantini, L. Balicas, J. Jiang, A. Gurevich, D.C. Larbalestier, R. Jin, A.S. Sefat, M.A. McGuire, B.C. Sales, D.K. Christen, D. Mandrus, *Appl. Phys. Lett*. **94,** 062511 (2009)

7. J. Karpinski, N.D. Zhigadlo, S. Katrych, Z. Bukowski, P. Moll, S. Weyeneth, H. Keller, R. Puzniak, M. Tortello, D. Daghero, R. Gonnelli, I. Maggio-Aprile, Y. Fasano, O. Fischer, B. Batlogg, *Physica C*, **469** 370 (2009)

8. N. Ni, S. L. Bud'ko, A. Kreyssig, S. Nandi, G. E. Rustan, A. I. Goldman, S. Gupta, J. D. Corbett, A. Kracher, P. C. Canfield, *Phys. Rev. B* **78**, 014507 (2008)

9. G. F. Chen, Z. Li, J. Dong, G. Li, W. Z. Hu, X. D. Zhang, X. H. Song, P. Zheng, N. L. Wang, J. L. Luo, *Phys. Rev. B* 78, 224512 (2008)

10. C. Senatore, R. Flükiger, M. Cantoni, G. Wu, R. H. Liu, and X. H. Chen, *Phys. Rev. B* **78**, 054514 (2008)

11. E.S. Otabe, M. Kiuchi, S. Kawai, Y. Morita, J. Ge, B. Ni, Z. Gao, L. Wang, Y. Qi, X. Zhang, Y. Ma, *Physica C* **469,** 1940 (2009)

12. S. Lee, J. Jiang, J. D. Weiss, C. M. Folkman, C. W. Bark, C. Tarantini, A. Xu, D. Abraimov, A. Polyanskii, C. T. Nelson, Y. Zhang, S. H. Baek, H. W. Jang, A. Yamamoto, F. Kametani, X. Q. Pan, E. E. Hellstrom, A. Gurevich, C. B. Eom, D. C. Larbalestier, *Cond-mat:* arXiv, 0907.3741 (2009)

13. Y. Matsumoto, J. Hombo, Y. Yamaguchi, T. Mitsunaga, *Materials Research Bulletin* **24**, 1469 (1989); Y. Matsumoto, J. Hombo, Y. Yamaguchi, *Materials*

*Research Bulletin* **24**, 1231 (1989)

14. F. Y. Chuang, D. J. Sue, and C. Y. Sun, *Materials Research Bulletin* **30**, 1309 (1995)

15. X.G. Zheng, H. Matsui, S. Tanaka, M. Suzuki, C.N. Xu, and K. Shobu, *Materials Research Bulletin* **33**, 1213 (1998)

16. Y. Matsumoto, J. Hombo, and Y. Yamaguchi, *Appl. Phys. Lett.* **56**, 1585(1990)

17. Yanwei Ma, Zhaoshun Gao, Lei Wang, Yanpeng Qi, Dongliang Wang, Xianping Zhang, *Chin. Phys. Lett.* **26**, 037401 (2009)

18. Yanpeng Qi, Xianping Zhang, Zhaoshun Gao, Zhiyu Zhang, Lei Wang, Dongliang Wang, Yanwei Ma, *Physica C* **469**, 717 (2009)

19. J. D. Moore, K. Morrison, K. A. Yates, A. D. Caplin, Y. Yeshurun, L. F. Cohen, J. M. Perkins, C. M. McGilvery, D. W. McComb, Z. A. Ren, J. Yang, W. Lu, X. L. Dong and Z. X. Zhao, *Supercond. Sci. Technol.* **21**, 092004 (2008)

20. F. Kametani, P. Li, D. Abraimov, A. A. Polyanskii, A. Yamamoto, J. Jiang, E. E. Hellstrom, A. Gurevich, D. C. Larbalestier, Z. A. Ren, J. Yang, X. L. Dong, W. Lu, Z. X. Zhao, *Cond-mat:* arXiv, 0907.4454 (2009)

21. L. Wang et. al Cond-mat: arXiv, 0911.3701 (2009)

**Captions**

Figure 1 X-ray diffraction patterns for the pure and Ag added $Sr_{0.6}K_{0.4}Fe_2As_2$ samples.

Figure 2 Temperature dependence of resistivity for the pure and Ag added samples; Inset: Temperature dependence of DC susceptibility of the pure (black squares) and 5 % Ag added (red circles) samples.

Figure 3 The upper critical field $H_{c2}$ and irreversibility field $H_{irr}$ as a function of temperature of various samples

Figure 4 $J_c$ yielded from the hysteresis of the bulks at 5 K and 20 K (inset).

Figure 5 Scanning electron micrographs of the pure (a), 5 % (b) and 20 % (c) Ag added samples; (d) a typical QBSD image on polished surface for the Ag added samples.

Figure 6 Transmission electron microscopy (TEM) images of the pure samples.

Figure 7 TEM images of the Ag added samples.

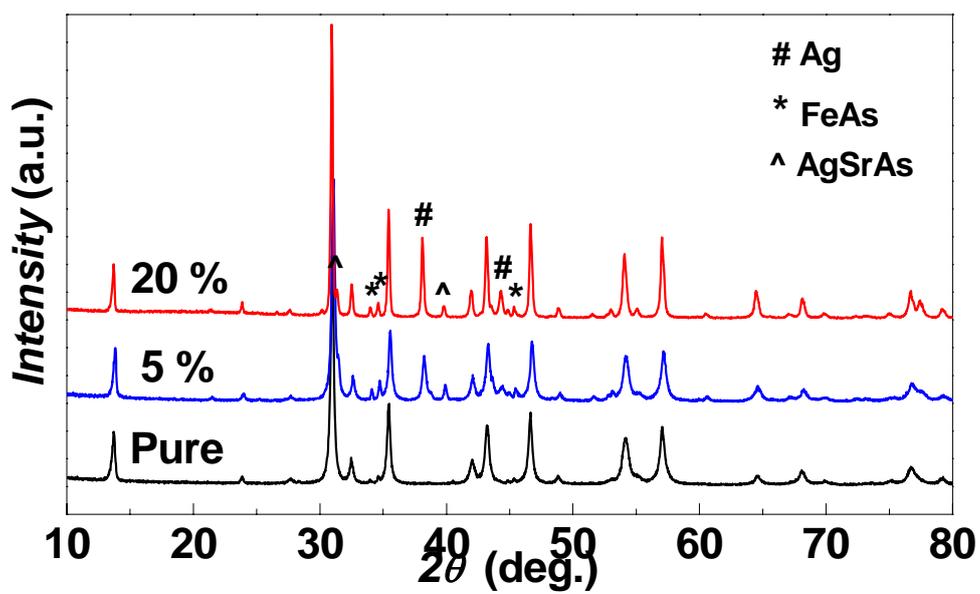

Figure 1 Wang et al.

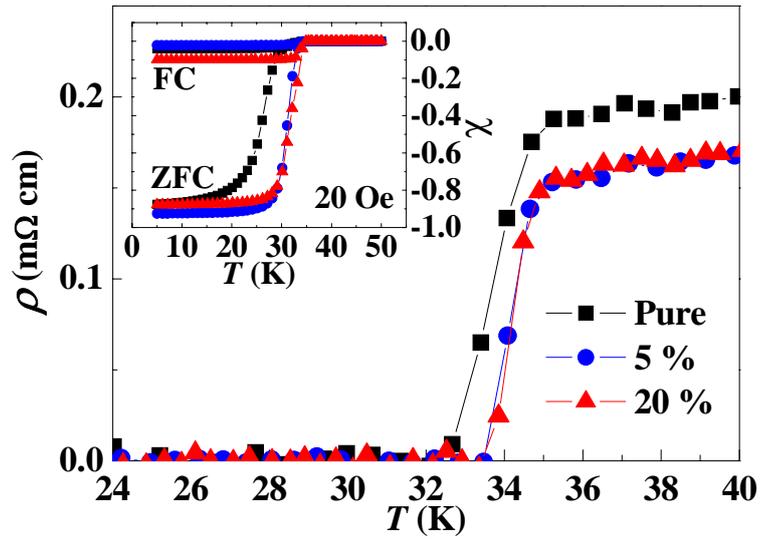

Figure 2 Wang et al.

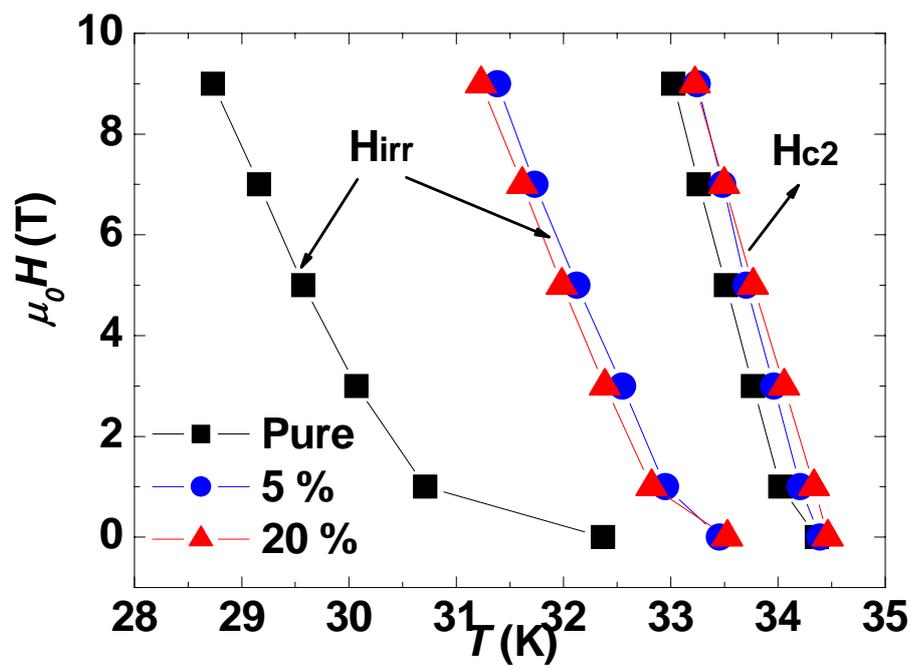

Figure 3 Wang et al.

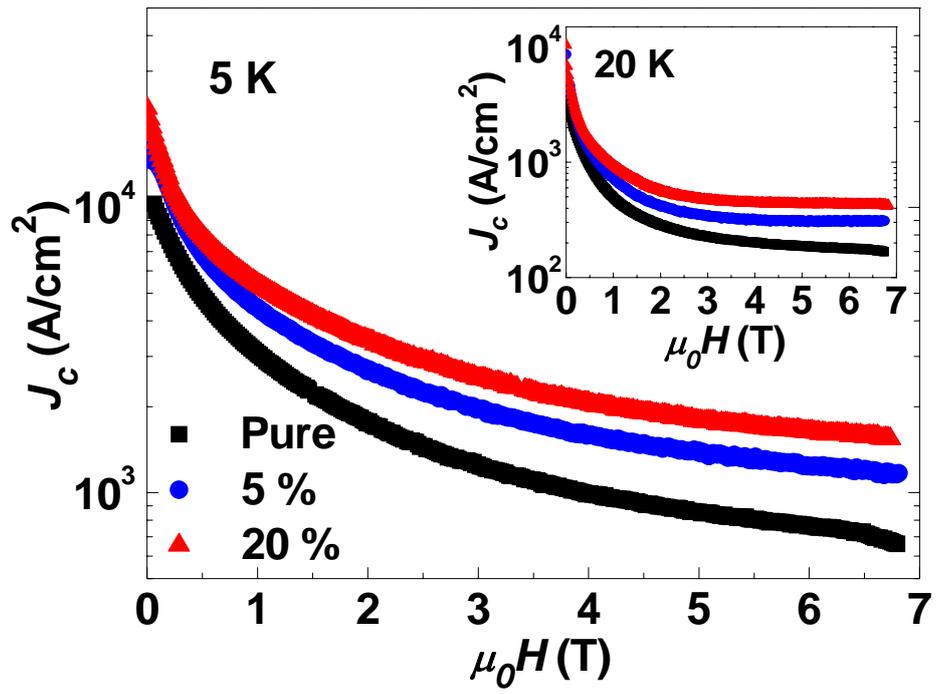

Figure 4 Wang et al.

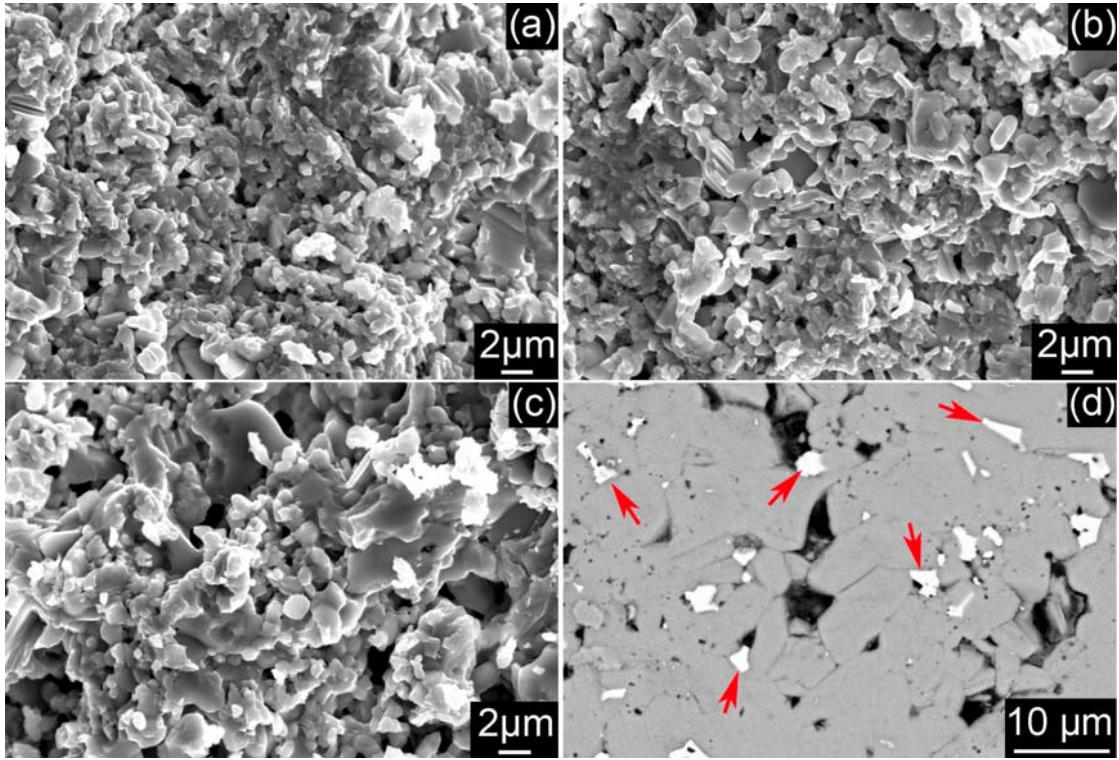

Figure 5 Wang et al.

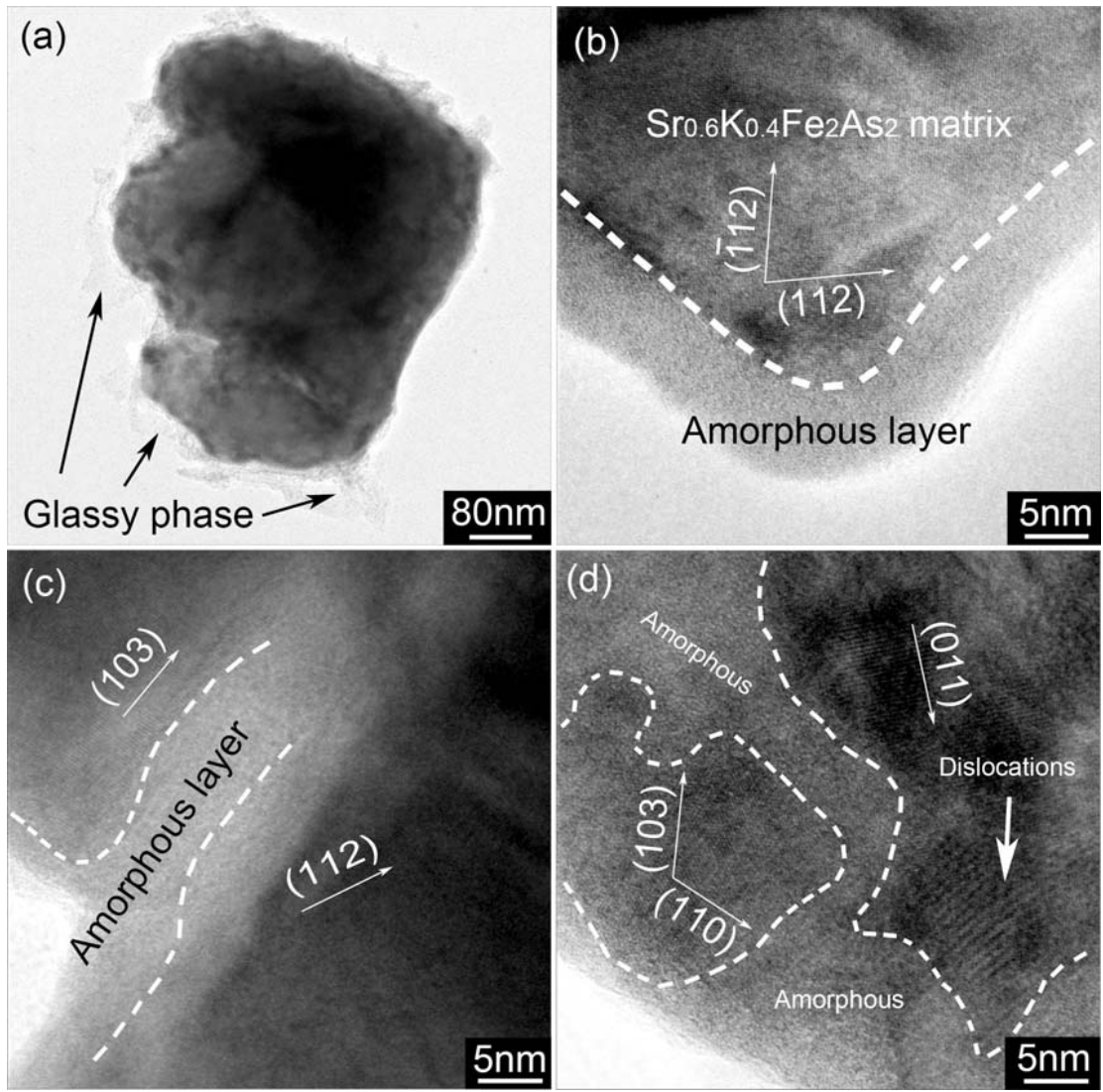

Figure 6 Wang et al.

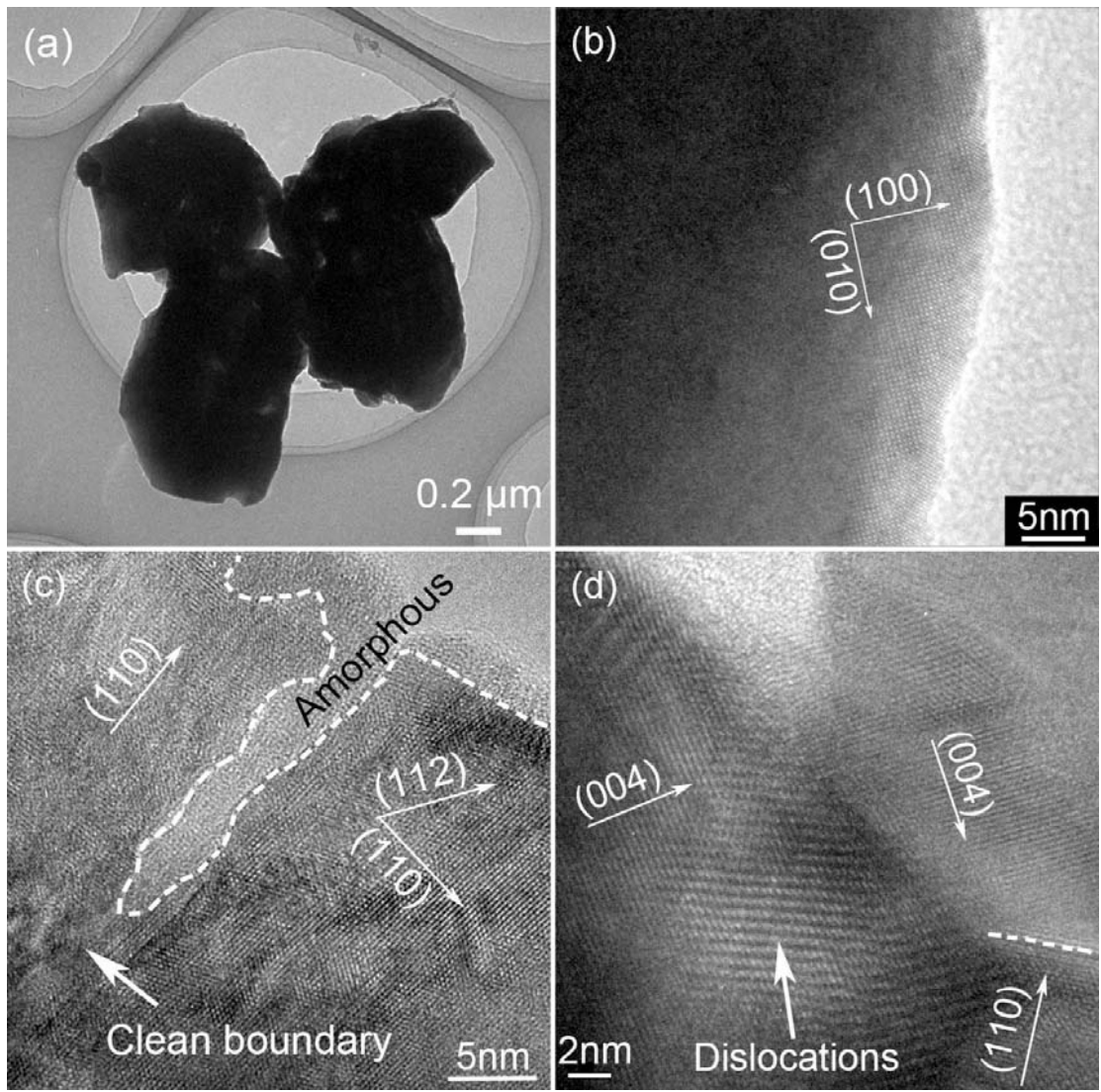

Figure 7 Wang et al.